\documentclass[useAMS, usenatbib]{mn2e}
\usepackage{times}
\usepackage{amsmath}
\usepackage{xspace}
\usepackage{psfig}
\usepackage{psfig}
\usepackage{defs}

\def\eq{equation}
\def\fig{Fig.}

\def\ie{{\it i.e.}}
\def\eg{{\it e.g.}}
\def\ltsima{$\; \buildrel < \over \sim \;$}
\def\simlt{\lower.5ex\hbox{\ltsima}}
\def\gtsima{$\; \buildrel > \over \sim \;$}
\def\simgt{\lower.5ex\hbox{\gtsima}}

\def\h2{H$_2$\xspace}

\def\Mpc{h$^{-1}$ Mpc\xspace}

\title[On the Origin of Dark Matter Cores in Dwarf Galaxies]{%
  On the Origin of Dark Matter Cores in Dwarf Galaxies}

\author[M. Ricotti \& M.~I. Wilkinson]
{Massimo Ricotti and Mark I. Wilkinson\\ 
Institute of Astronomy, Madingley Road, Cambridge CB3 0HA\\
ricotti@ast.cam.ac.uk}
\date{Accepted ---. Received ---; in original form 19 August 2003}
\pagerange{\pageref{firstpage}--\pageref{lastpage}}
\pubyear{2003}
\begin{document}
\maketitle
\label{firstpage}

\begin{abstract}
  In this paper, we study the dynamical stability and time evolution
  of the central dark matter cores of low-mass (about
  $10^8-10^9$M$_\odot$) galactic haloes found in recent cold dark
  matter simulations at high redshift. From those simulations we
  extract three haloes, assembled by hierarchical merging, that at
  redshift $z \simgt 10$ display a core and we evolve them without
  further merging to low redshift using direct $N$-body integration.
  The central core in the dark matter profile is found to be
  dynamically stable: it survives for many crossing times without
  evolution into a cusp. This result supports the claim that the mass
  dependence of the central dark matter profiles of simulated haloes
  is a direct consequence of the power spectrum of primordial density
  fluctuations. In addition, we show that the simulated dark matter
  profiles, if they evolved in isolation, are consistent with the
  observed velocity dispersion profile of stars in the inner parts of
  the Draco dwarf spheroidal galaxy. Simple scaling arguments are
  reviewed which explain the evolution of the concentration parameter
  with redshift. We also review some arguments used to derive the
  logarithmic slope of the inner and outer density profile.
\end{abstract}

\section{Introduction}

Most of the mass in our universe is thought to be in the form of an
unknown dark and collisionless matter that interacts with itself and
with normal matter (baryons) only gravitationally. The formation of
galaxies and the large scale structures in our universe is driven by
the gravitational forces exerted by this dark matter (DM). On large
scales the agreement of observations with models where the dark matter
can cluster on all mass scales (cold dark matter: CDM) is very good.
In this cosmological model, small-mass structures collapse first while
larger galaxies and clusters form later by accretion and merging of
smaller subclumps.  In recent years, it has become possible to
investigate the nature of DM in more detail both because of
improvements in numerical simulations and because it has been realised
that the Local Group of galaxies is a perfect observational laboratory
in which to test the predictions of DM simulations.

Observations of rotation curves in nearby dwarf galaxies and low
surface brightness (LSB) galaxies generally show that their DM density
profile has a flat inner core
\citep{vandenBosch:00,deBlok:01,deBlok:02, BorrielloS:01,SalucciB:00}.
This appears to be in disagreement with the results of numerical
simulations of CDM haloes. Using $N$-body simulations, several groups
find that halo density profiles generally have an inner cusp with
slope $\alpha =1-1.5$ and, most remarkably, that the mean shape of the
profile is universal: it is the same from dwarf galaxies to clusters
at any redshift (\eg, \citealp{Navarro:97, Moore:99} but see
\citealp{SyerW:98, Kravtsov:98, SubramanianCO:00, JingS:00}).  A large
number of solutions to resolve the disagreement between observations
and theory have been proposed. They can be divided into two main
categories: (i) feedback effects of the baryons and stars on the DM
density profile and (ii) modifications of the physical properties
(\eg, scattering cross section, temperature) of the DM particles
\citep[\eg,][]{BodeO:01, Spergel:00}. The first class of solutions
does not seem to be very effective in removing cusps and while the second
has achieved moderate success in explaining the absence of a large
population of dwarf haloes around the Milky Way, this comes at the
expense of introducing additional inconsistencies with the standard
cosmological model.

The disagreement between observations and CDM simulations mainly
concerns low-mass haloes and small length scales. In hierarchical
scenarios, small mass haloes are mostly assembled at high-redshift. On
the other hand $N$-body simulations have focused on studying the
formation of Milky-Way type haloes or clusters ($10^{11} {\rm M}_\odot
\simlt M \simlt 10^{14}$ M$_\odot$) at low-redshift ($0 \simlt z_{\rm
vir} \simlt 3$). Recently, \citep[][hereinafter RO3]{Ricotti:03}
have used $N$-body simulations to study the DM profiles of haloes with
masses between $10^8$ M$_\odot$ and $10^{15}$ M$_\odot$, comparing
them at their characteristic redshift of formation ($0 \simlt z_{\rm
vir} \simlt 10$).  This is done using simulations with different box
sizes and comparing the DM profiles of the haloes at different
redshifts, when the most massive haloes in the simulations are
composed of about the same number of DM particles. The results show
that the density profile of DM haloes is not universal, presenting
cores in dwarf galaxies and steeper cusps in clusters. This result
could be a resolution of the conflict between the results of CDM
simulations and the observed dwarf galaxy rotation curves, if we
assume that the profile is stable and remains flat from $z=10$ (when
the simulation is halted because of the small volume of the box) to
low redshift.

It should be mentioned that in a recent work, \cite{Colin:03} have
analysed simulations with similar resolution and box sizes as those in
R03 and find no evidence for a trend of the inner slope with the mass
of the dark halo. This disagreement may be due to differences in their
analysis of the simulations, as they have analysed the slopes of the
haloes only at redshift $z=3$ for all their simulations, irrespective
of box size.  It is well known that the smaller the box size the
earlier a simulation ceases to be a representative volume of the
universe. In a simulation with 1\Mpc (32 \Mpc) box size at $z \sim 10$
($z=3$), modes with wavelength comparable to the size of the box
become nonlinear and, for instance, the halo mass function becomes
unreliable.  \cite{Colin:03} conclude that the results of their
simulations are in disagreement with R03 because the simulations are
identical; however, we note that they have analysed their haloes at
redshift $z=3$ while the haloes in R03 were analysed at redshift
$z=10$. As a consequence of the additional evolution from $z=10$ to
$z=3$, their haloes are about ten times more massive. It is therefore
likely that the disagreement is related to their attempt to analyse
haloes at low redshift in a small cosmological volume.  Indeed, R03
also finds that the cores observed in the ten most massive haloes at
$z < 10$ eventually evolve to cusps.  Any possible trend of the dark
halo inner slopes with halo mass can easily be confused with a mass
dependence of the concentration parameter. In order to demonstrate
that a trend exists, particular care must be taken to minimise any
possible systematic errors, for example, by comparing haloes in
simulations with different box sizes only when the most massive haloes
have comparable number of particles (R03). More recently,
\cite{CenO:04} have confirmed the results found by R03; in addition,
they identify a redshift dependence of the typical halo profile.

In addition to the disagreement on the value of the inner slope
(\cite{Power:03} find $\alpha=1$ while $\alpha=1.5$ is advocated by
\cite{Moore:99}) other groups have also found deviations of the DM
profile from a ``universal shape''. These deviations manifest
themselves as dependences of the profile shape on the mass,
environment, cosmology or merging history \citep{SyerW:98,
Kravtsov:98, SubramanianCO:00, JingS:00, Jing:00}.  In recent years,
it has become possible to observe the DM profile of haloes at the two
extremes of the halo mass function: in dwarf spheroidal galaxies with
$M \sim 10^8$ M$_\odot$ and in clusters with $M \sim 10^{15}$
M$_\odot$. In \S~\ref{sec:intob} we summarise observational evidence
that appears to confirm the results found by R03, albeit with
significant uncertainty.

The simulation results in R03 strongly suggest that the mass function
of the accreting satellites determines the inner slope of the DM
profile. However, the results might also be interpreted as a trend
with dynamical state in the assembly process of haloes of different
mass.  In this paper we study the dynamical stability of the flat
profiles of dwarf-size haloes at $z=10$ by evolving them from their
redshift of formation to lower redshift without further merging. We
extract a few haloes at $z=10$ and we evolve them in isolation using a
direct $N$-body integration code (see \S~\ref{sec:sim} for details on
the method). In \S~\ref{sec:res} we show that the profiles are
dynamically stable and retain an almost flat core.  As shown in
\S~\ref{sec:dsph} the DM profile is consistent with the observed
velocity dispersion of stars in the inner regions of the Draco dSph.
Simple scaling arguments are given in \S~\ref{sec:scal} to understand
how the concentration parameter evolves with redshift and what
determines the slope of the outer profile. We summarise the results in
\S~\ref{sec:con}.

\section{Non-universal DM profile: Observations}\label{sec:intob}

There is growing observational evidence that the DM density profile in
the central regions of low-mass galaxy haloes is not universal and
that cored profiles are favoured over cuspy halo models.  Despite the
on-going debate about the correct interpretation of the gas rotation
curves of low-mass disk galaxies, it is clear that these galaxies
display a wide range of inner halo profiles. What is also apparent is
that the data generally favour logarithmic density slopes close to
$0.2$ \citep[\eg,][]{deBlok:03}. The gas-free Local Group dwarf
spheroidal galaxies provide a cleaner test-bed for DM models as they
are often completely DM dominated at all radii and it is therefore
possible to measure their DM content by treating their baryonic
content as a massless tracer population. The recent identification of
a kinematically cold stellar substructure in the Ursa Minor dSph
\citep{Kleyna:03} strongly suggests that the dark halo of that galaxy
has a central core. Using $N$-body simulations \cite{Kleyna:03} showed
that this cold structure would survive for less than $1$ Gyr if the DM
core were cusped. Only if the DM core has a uniform density core can the
cold substructure survive for a Hubble time. Additionally,
\cite{Magorrian:03} found $\alpha = 0.55^{+0.37}_{-0.33}$ for the
Draco dSph.

Gravitational lensing observations show that the inner slope of the DM
profile in clusters is $1 \simlt \alpha \simlt 1.5$.
\cite{Dahle:03} studied six massive clusters with masses $M \sim
10^{15}$ M$_\odot$ at $z=0.3$.
They find $\alpha=1.4^{+0.2}_{-0.1}$ and $c=1.7^{+0.9}_{-1.3}$ at the
68 \% confidence level.  \cite{Smith:01} finds $\alpha = 1.3$ for the
lensing cluster A383 that has a mass $M \sim 10^{14}$ M$_\odot$
($z=0.188$).  The lensing cluster MS 2137-2353 has been studied
independently by two groups. Their conclusions are quite different.
\cite{Gavazzi:03} find $0.7 \le \alpha \le 1.2$, but they note that if the
identification of a fifth lensed image is confirmed by HST
observations, an isothermal profile with a flat core fits the
data better.
\cite{Sand:02} find a best-fit $\alpha = 0.35$ and $\alpha < 0.9$ at
99 \% confidence level for the same cluster. These results are
extended and strengthened by the analysis of six clusters presented in
\cite{Sand:04}. X-ray studies based on Chandra observations find that
the inner slope of the DM profile is consistent with $\alpha=1$
\citep{Arabadjis:02}.  \cite{Lewis:03} use a generalised profile to
fit their observations and find $\alpha \approx 1.2 \pm 0.04$ for the
central slope of A2029.

Thus there is some observational evidence for a dependence on halo
mass of the inner slope of dark matter haloes, and hence for the
non-universality of the dark matter profile, in agreement with the
findings of R03.

\section{Numerical Method}\label{sec:sim}
We extract three haloes (HALO~1, 3 and 7) from the simulation with
$L_{\rm box}=1$ \Mpc in R03 at $z=10$ and we evolve each
of them as an isolated $N$-body system using a direct
$N$-body code as explained in the next subsection.  We can neglect the
Hubble expansion since for $r \ll V_{\rm c}(r)/H(z) \approx 10$ kpc,
where $V_{\rm c}$ is the halo circular velocity, the effect on the
inner profile is negligible.  The location of the dark matter core
radius, $r_{\rm s} \sim 200$ pc, is not affected by the Hubble
expansion.

\subsection{$N$-body simulations}

We follow the evolution of the low mass haloes using the direct
$N$-body integration code NBODY4~\citep{Aarseth:99}.  In order to
avoid spurious two-body interactions (that are non-physical for dark
matter particles) we have adopted a standard force smoothing at small
particle separation with the same Plummer softening parameter,
$\epsilon \approx 11$ pc (physical) as that used in R03.  We have also
run the same simulations without force smoothing to investigate the
possible importance of spurious two-body effects, finding that the
profiles at early times are very similar. Even at later times there
are only negligible differences between the evolved profiles in the
two sets of runs (softened and un-softened).
Initially, the three simulations performed contain, respectively,
$54341$, $46935$ and $26565$ particles of mass $7050$ M$_\odot$ and
their evolution is followed for about $5$ Gyr ($12$ Gyr for the
un-softened runs). The simulations were performed on GRAPE-6
special-purpose computer boards \citep{Makino:97} at the American
Museum of Natural History, New York and at the Institute of Astronomy,
Cambridge.  At regular intervals during the integration the positions
and velocities of all particles in the simulation were recorded and
the density profiles calculated.  The density profiles are calculated
as in R03.  The centre of the halo is calculated iteratively for
particles within a given radius. The radius is reduced until the
centres calculated using the centre of mass, density maximum and the
minimum of the gravitational potential, converge. The results converge
within $2\epsilon$, where $\epsilon \approx 11$ pc is the Plummer
softening parameter.

\section{Results}\label{sec:res}

The main aim of this paper is to investigate whether the profiles
found by R03 are in equilibrium.  In order to check the stability of
the haloes, however, we do not need to evolve the halo for 12 Gyr (the
time from $z=10$ to $z=0$) but for a much shorter time scale: namely
the crossing time $t_{\rm cr} \sim 0.5-1$ Gyr. On this time scale
two-body relaxation can be safely ignored since $t_{\rm cr} < t_{\rm
  relax}$, where $t_{\rm relax}$ is the two-body relaxation time
scale.

\begin{figure}
\centerline{\psfig{figure=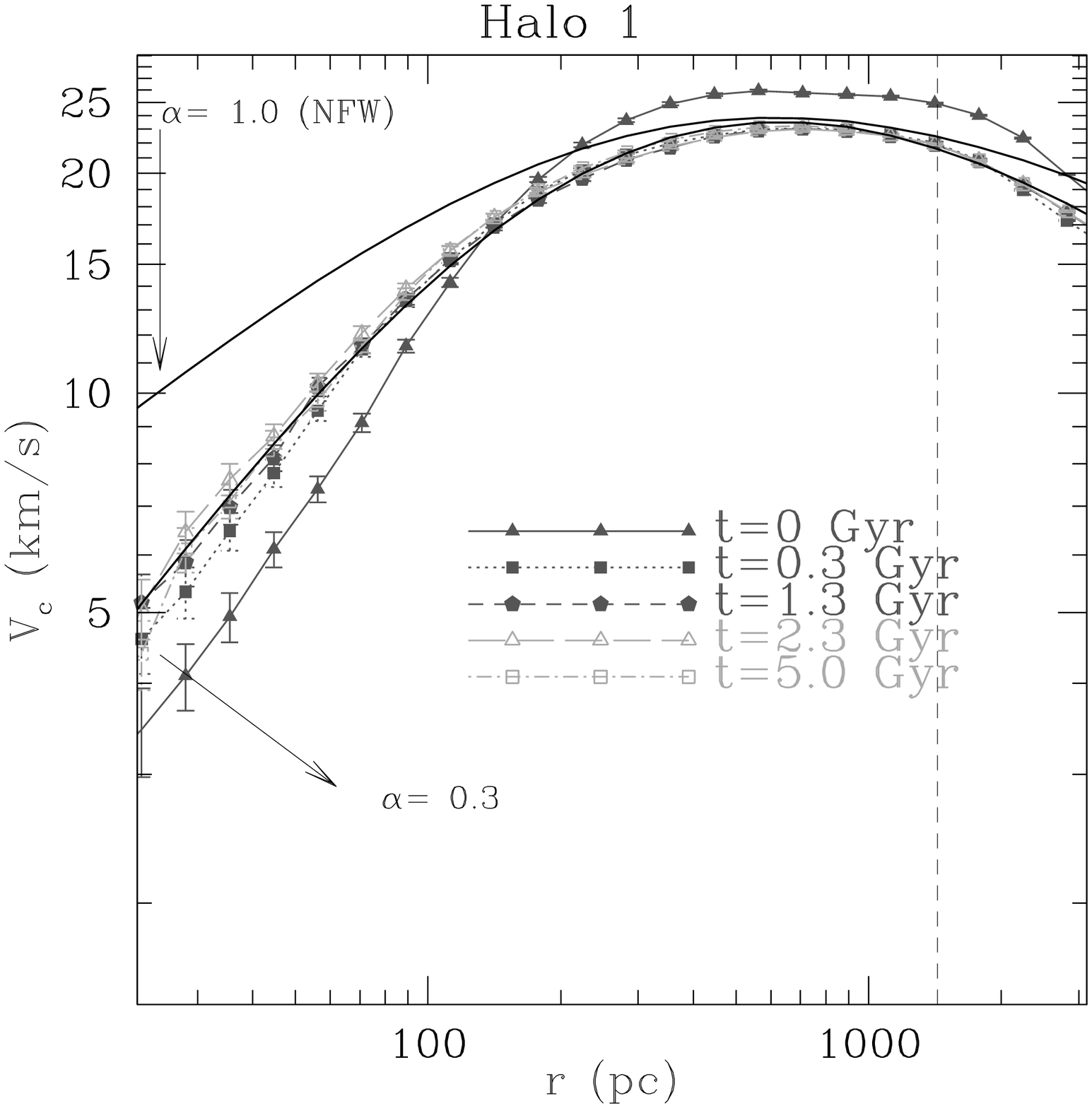,width=7.3cm}}
\centerline{\psfig{figure=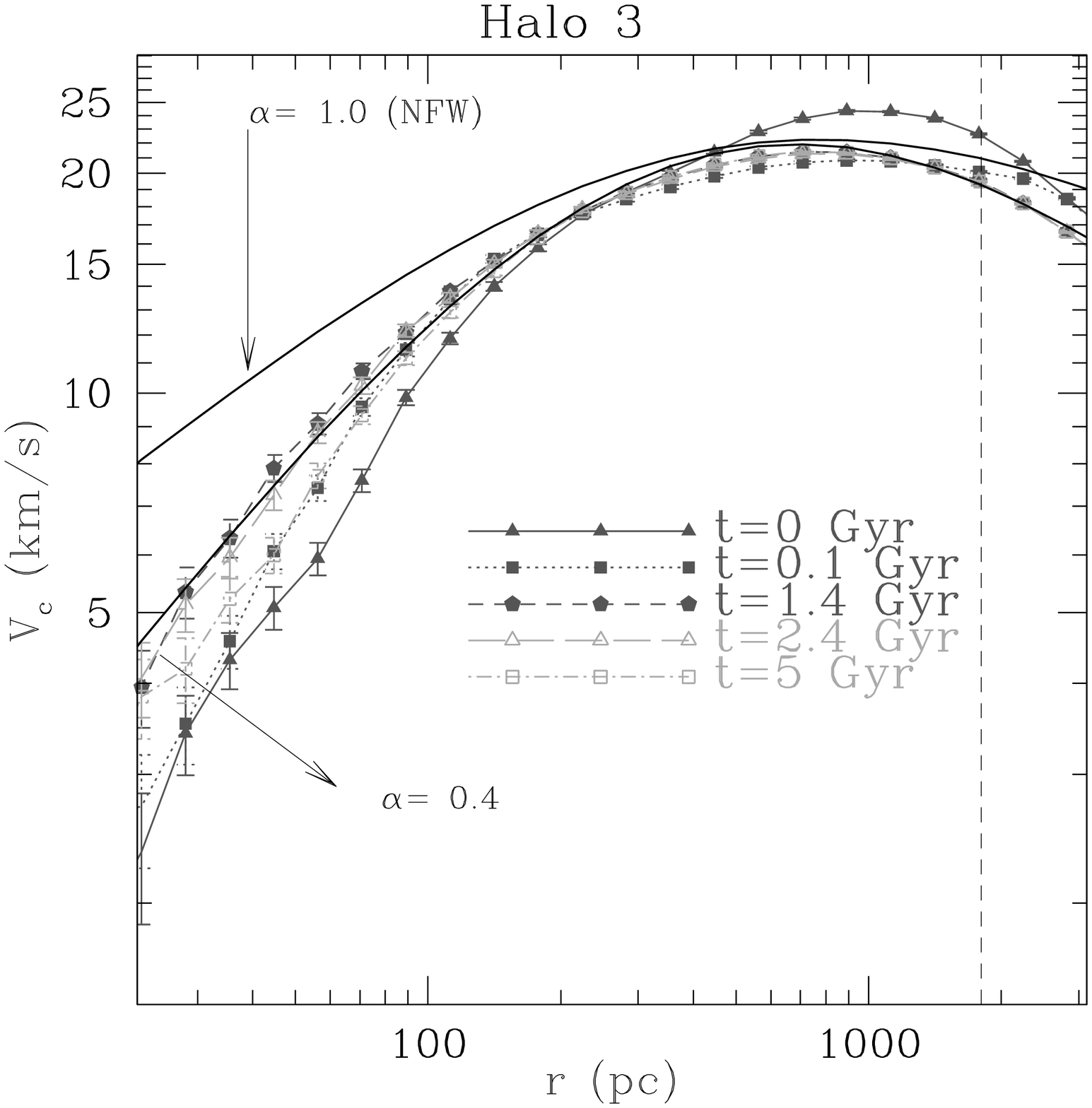,width=7.3cm}}
\centerline{\psfig{figure=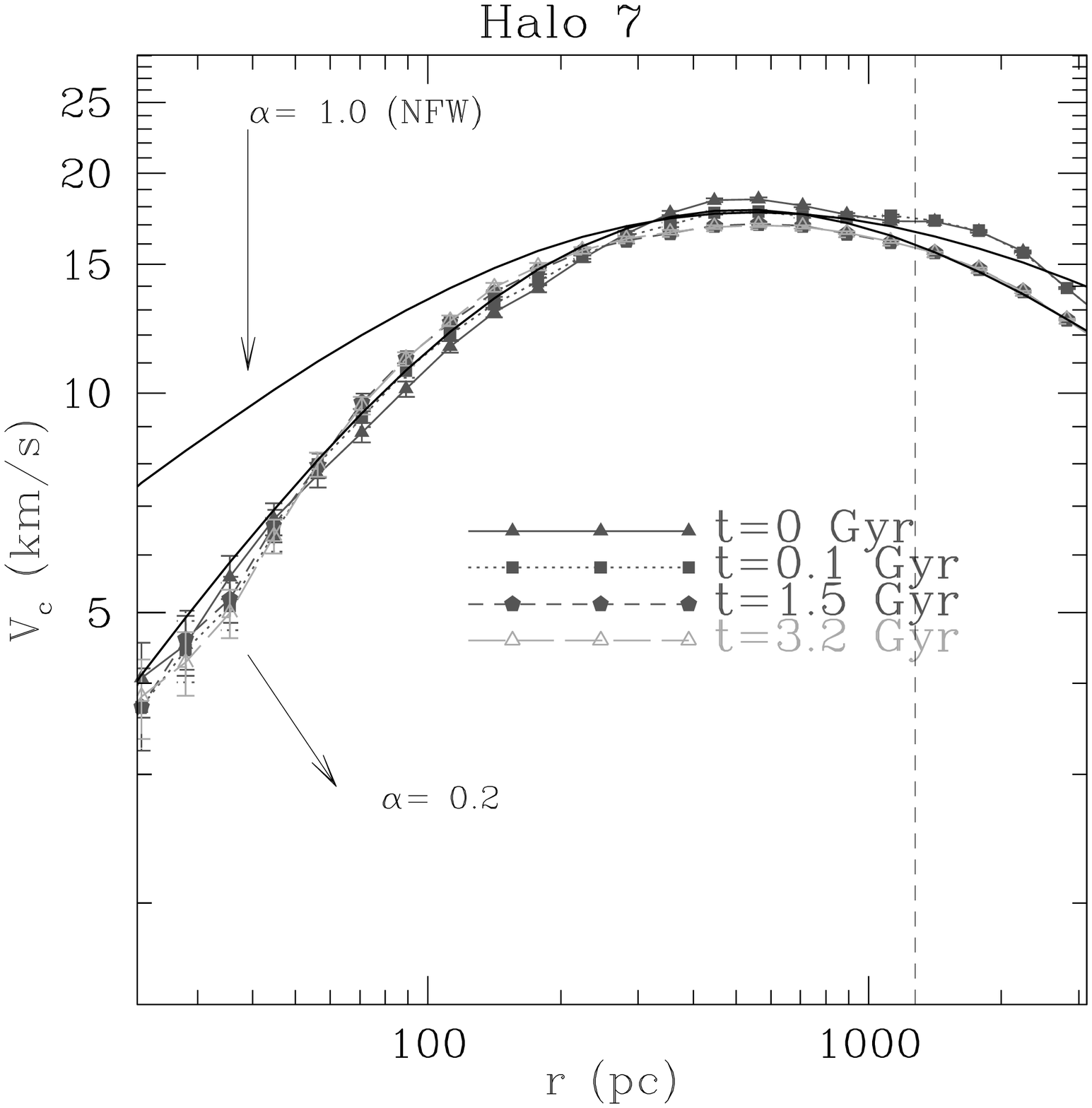,width=7.3cm}}
\caption{\label{fig:halo} The circular velocity $V_{\rm c}$ of 3 haloes with
  $M \sim 4 \times 10^8$ M$_\odot$ virialised at $z \sim 10$ and
  evolved in isolation for a time $t \gg t_{\rm cr}$. The circular
  velocity profiles are shown at the times after $z=10$ shown by the
  labels. Three haloes are shown: HALO~1 ($M \approx 3.8 \times 10^8$
  M$_\odot$), HALO~3 ($M \approx 3.3 \times 10^8$ M$_\odot$) and
  HALO~7 ($M \approx 1.9 \times 10^8$ M$_\odot$) from the $L_{\rm
    box}=1$ \Mpc simulation in R03. The dashed line shows
  the location of the virial radius when the haloes formed at $z \simgt
  10$. The lower limits on the x axis correspond to twice the
  force softening length.}
\end{figure}

\fig~\ref{fig:halo} shows the the circular velocity $V_{\rm
  c}=(GM(r)/r)^{1/2}$ of the three haloes with $M \simeq 2-4 \times
10^8$ M$_\odot$ virialised at $z \simgt 10$ and evolved in isolation
for about $5$ Gyr. The profiles are shown at the times $t$ after
$z=10$ indicated by the labels. The vertical dashed lines show the
location of the virial radius when the haloes formed at $z=10$. Note
that at $z=0$ the halo virial radius (defined as the radius where the
halo mean density is about $200$ times the IGM mean density) will be
about ten times larger (see \S~\ref{ssec:inner}) but a fraction
$\simgt 20$\% of the halo mass will reside in the inner 1 kpc. If
instead the halo is accreted by a larger galaxy the outer envelope of
the halo will be lost and the mass and radius might be reduced to the
values at the epoch of formation ($z=10$). The profiles shown
here are reliable in the range of radii $r> 2\epsilon=22$ pc, where
$\epsilon$ is the force smoothing length and $r \simlt 1$ kpc.  At
larger radii, the Hubble expansion and the mean density of the
intergalactic medium (IGM) should be included to determine the
evolution of the outer density profile. We will return to this issue
in \S~\ref{ssec:inner}.

As the figures show, after about $0.1$ Gyr, the profiles settle into a
stable configuration and the central regions of the haloes (within
about $1$ kpc) do not evolve significantly during the next 5 Gyrs. The
final density profiles of the three haloes are very similar to each
other and have almost flat cores with inner slopes in the range
$\alpha \sim 0-0.4$. After $t \sim 5$ Gyrs two-body relaxation may
start to be important in driving the evolution of a mildly cusped core
into a halo with a flatter core \citep[see \eg,][]{Hayashi:03,
  Jin:04}. This can be observed in HALO~3 that at $t \sim 1$ Gyrs has
settled to an equilibrium slope with $\alpha \sim 0.4$ and at $t>5$
Gyrs has evolved to an almost flat core ($\alpha \sim 0$).

The two-body relaxation time is $t_{\rm relax} \propto
\sigma^3/(m_{\rm p} \rho)$, where $\sigma$ is the velocity dispersion
of the particles, $m_{\rm p}$ is their mass and $\rho$ is their
density. Taking $\sigma \sim V_{\rm c}$ and the core density for
$\rho$ we have,
\begin{equation*}
t_{\rm relax}=(2.6~{\rm Gyr}) \left({M \over 10^8 M_\odot} \right)^{1/2}
  \left({ 7050 M_\odot \over m_{\rm p}}\right) \left({r_{\rm core} \over
  200 {\rm pc}}\right)^{3/2}
\label{eq:trel}
\end{equation*}
Since both the density profile at $t=0$ and the velocity dispersion
$\sigma$ are approximately constant with radius, $t_{\rm relax}$ does
not depend strongly on the radius.  For the haloes in our simulations
the estimated two-body relaxation time is $t_{\rm relax} \sim 5$ Gyr,
consistent with the observed evolution of HALO~3. The equilibrium
profiles of HALO~1 and HALO~7 (after $t\sim 1$ Gyrs) are almost flat
and therefore the cores in these haloes do not arise from any two-body
effects in the present work.  Moreover, since the Hubble time at
$z=10$ is ten times shorter than $t_{\rm relax}\sim 5$ Gyr, two-body
relaxation effects are not important even in the R03 simulations from
which the initial conditions where taken (see R03 for extensive
discussion of this issue).

Even the later evolution (at $t\sim 12$ Gyrs) is not strongly affected
by two-body relaxation, in the sense that the profile does not evolve
significantly towards a post core-collapse cusp.  Indeed one would
expect that relaxation would turn our initially cored profiles into
cusped profiles (via core collapse) on time scales $t \sim 17 t_{\rm
  relax} \sim 85$ Gyr for our haloes~\citep{Takahashi:95}.  We do not
see the effects of core collapse in our systems during the first 12
Gyr even in the simulations that do not include force softening.

\section{Comparison with observations}\label{sec:dsph}

\begin{figure}
\centerline{\psfig{figure=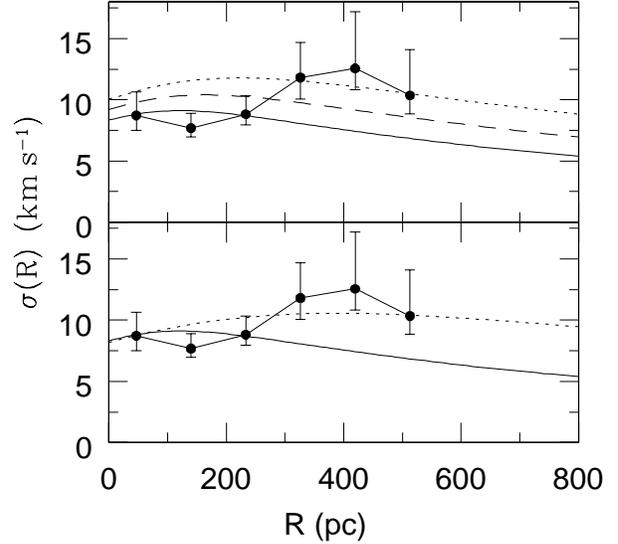,width=8cm}}
\caption{Comparison of the projected velocity dispersion profile for a
  S\'ersic light distribution embedded in the dark matter distribution
  of HALO~1 (solid line) with the observed profile for the Draco dSph
  (points with errorbars). In the top panel we also show two synthetic
  scaled versions of HALO~1 ($M_{\rm dm}\simeq 10^{8}$ M$_\odot$) with
  the same core density but with twice its mass (dashed line) and four
  times its mass (dotted line).  The dotted line in the bottom panel
  shows a halo eight times more massive than HALO~1 but with half its
  core density (\ie, virialised at $z<10$). Observed data taken from
  \protect\cite{Kleyna:02}.}
\label{fig:Draco_disp}
\end{figure}

In recent years, the volume of kinematic data available for Local
Group dSph galaxies has increased
significantly~\citep[\eg,][]{Mateo:97,Kleyna:02,Kleyna:03}. These data
make it possible for the first time to map the velocity dispersion
profiles of these galaxies and have confirmed that in at least one
case, the Draco dSph, the high central velocity dispersion persists to
large radii~\citep{Kleyna:01}. It is instructive to compare the
observable dispersion profile expected for a dwarf galaxy whose halo
mass distribution is matched by one of those simulated in the previous
section.  We assume that the projected light distribution in the dwarf
follows a S\'ersic profile
\begin{equation}
\Sigma_{\rm s} (R) = \Sigma_0 \exp\left[ -\left( R/R_{\rm s}\right)^{n_{\rm s}}\right],
\end{equation}
where $R_{\rm s} = 177$ pc and $n_{\rm s}=1.2$ which is a good fit to
the inner surface brightness profile of Draco
\citep{Klessen:03}. If we further assume that the stellar velocity
distribution is everywhere isotropic then integration of the Jeans
equations yields the expected (projected) velocity dispersion as a
function of radius
\begin{equation}
\sigma^2(R)={\int_R^{\infty} {dx \over x} \rho_{\rm s}(x)
V_{\rm c}^2(x)(x^2-R^2)^{1/2} \over \int_R^{\infty} dx \rho_s(x) x (x^2-R^2)^{-1/2}},
\label{eq:jeans}
\end{equation}
where $\rho_{\rm s}$ is the density of the light.
Fig.~\ref{fig:Draco_disp} compares the dispersion profile for HALO~1
at $z=0$ (solid line) with the observed dispersion profile for
Draco~\citep{Kleyna:02}. The other lines show the dispersion profiles
using synthetic scaled versions of HALO~1 (that has a final mass of
$M_{\rm dm}\simeq 10^{8}$ M$_\odot$). In the top panel we show two
haloes with the same core density as HALO~1 but with twice its mass
(dashed line) and four times its mass (dotted line).  The dotted line
in the bottom panel shows a halo eight times more massive than HALO~1
but with half its core density. This profile provides a better fit to
the overall shape of the dispersion profile. Haloes with total mass
$M_{\rm dm} \sim 2-8 \times 10^{8}$ M$_\odot$ provide reasonable fits
to the observed profile.
We note that~\cite{Stoehr:02} also found haloes in their simulations
which were able to match Draco's dispersion profile.  However, given
that the resolution limit (in physical units) of those simulations was
comparable to the edge of the light distribution in Draco it is not
clear that their agreement with the observed dispersion profile is
physically significant.

\section{Scaling arguments}\label{sec:scal}
In this section we present a short review of previous semi-analytic
arguments in order to illustrate the physical mechanisms that
determine the shapes of dark matter haloes. Using very simple
arguments, we attempt to clarify the links between our numerical
results and theory.  Using $N$-body simulations R03 have shown that
the density profile
\begin{equation}
\rho_{\rm dm} \propto {1 \over X^\alpha(1+X)^{\gamma-\alpha}}
\label{eq:rho}
\end{equation}
with $X=c_\Delta R/R_\Delta$, $\alpha =(9+3n)/(5+n) \approx
1.3+(M_{\rm dm,14}^{1/6}-1)/(M_{\rm dm,14}^{1/6}+1)$ and constant
concentration parameter $c_\Delta \sim 7$ provides a good fit to all
recently-virialised DM haloes from dwarf galaxies to clusters. Here,
$n$ is the effective spectral index of the initial power spectrum of
density perturbations, $R_\Delta(M_{\rm dm})$ is the virial radius and
$M_{\rm dm,14}=M_{\rm dm}/(3 \times 10^{14}$ M$_\odot$), where $M_{\rm
  dm}$ is the halo mass.

The relationship between the inner slope $\alpha$ and $n$ can be
understood using simple scaling arguments (see \S~\ref{ssec:inner}).
In the present paper, the haloes are evolved from $z=10$ to low
redshift in a zero-density universe, neglecting the Hubble expansion.
This simplification prevents us from knowing the evolution with
redshift of the concentration parameter and the outer profile of the
dark halo. In \S~\ref{ssec:outer} we review some simple arguments
whose aim is to clarify the physical processes that determine the
evolution of the concentration parameter and the outer slope of the
density profile from the time of halo virialisation to $z=0$.

\subsection{Inner slope}\label{ssec:inner}
In this section, following \cite{Peebles:74} and \cite{Hoffman:85}, we
review the arguments used to derive the inner profile of DM haloes
under the simplistic assumption that each sub-halo (here represented
as a step function in density) survives undisrupted when incorporated
in a larger halo. The result that we obtain agrees remarkably well
with the simulations in R03.

Assuming an Einstein de-Sitter universe, linear perturbations
$\delta_k$ grow as $\delta_k(t) \propto a(t) \propto t^{2/3}$, where
$a$ is the scale factor of the universe and $t$ is the age of the
Universe. The variance $\sigma^2$ in the amplitude of the density
perturbations varies with perturbation mass $M$ and time $t$ as
$\sigma^2(M,t) \propto t^{4/3} M^{-(3+n)/3}$.  Here $n$ is the
effective spectral index of the initial power spectrum of density
perturbations $P(k) \propto k^n$. When $\sigma^2(M,t)$ reaches
approximately unity, the associated perturbations become non-linear:
this occurs at time $t_{\rm vir}(M)$.  Hence, $t_{\rm vir} \propto
M^{(n+3)/4}$: small masses become non-linear at earlier times if
$n>-3$. The halo sizes as a function of mass are given by $R(M) =
a(t_{\rm vir}) r_{\rm co} \propto M^{(n+5)/6}$, where the comoving
radius is $r_{\rm co} \propto M^{1/3}$, and the mean densities of the
haloes are $\rho(M) \propto a(t_{\rm vir})^{-3} \propto M^{-(3+n)/2}$.
The profile of a halo formed by the merger of many such haloes is then
given by the envelope that contains all the merged haloes piled on
each other like ``Russian dolls''. Smaller, higher-density haloes will
reside in the centre while the more extended, lower-density haloes
dominate the profile at large radii. Combining the expressions for
$R(M)$ and $\rho(M)$ we find $\rho(R) \propto R^{\alpha}$, where
$\alpha = (9+3n)/(5+n)$. The effective $n$ in the 1 \Mpc simulations
of R03 is $-2.6 \pm 0.1$ corresponding to $\alpha=0.4-0.6$.

\subsection{Halo concentration and outer slope}\label{ssec:outer}

The post-virialisation evolution of {\em isolated} dark haloes can be
understood (to a first approximation) using simple arguments
partially based on the theory of secondary infall developed by
\cite{GunnG:72} and \cite{Gott:75}.
\begin{figure}
\centerline{\psfig{figure=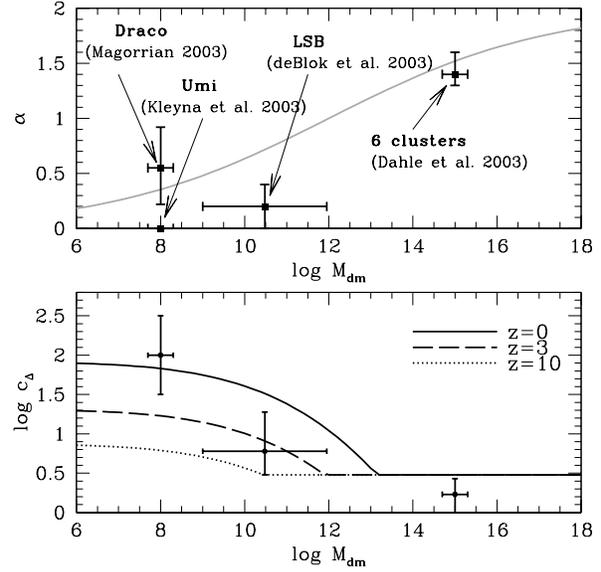,width=8cm}}
\caption{\label{fig:cm} (top) Slope $\alpha$ of the inner DM profile
  as a function of halo mass, $M_{\rm dm}$ (solid curve). The points
  with error bars are based on observations (see \S~\ref{sec:intob}).
  (bottom) Concentration parameter, $c_\Delta$, of the generalised
  profile at $z=0$ (solid line), $z=3$ (dashed line) and $z=10$
  (dotted line) as a function of halo mass. The points with errorbars
  are the concentrations at $z=0$ for the same haloes as in the top
  panel.  We estimate the virial radius of dSphs assuming that the DM
  density declines as $\rho \propto R^{-3}$ outside the stellar tidal
  radius.  Note that Sand et al. (2004) find an almost flat slope of
  the inner density profile in clusters, in disagreement with Dahle et
  al.  (2003) and others.}
\end{figure}


Particles that decouple from the Hubble flow and become
gravitationally bound are found, by definition, at radii where the
escape velocity is smaller than the Hubble expansion (\ie, $v_{\rm
  esc} \le v_{\rm Hubble}=H(z) R_{\rm halo}$, where $H(z)$ is the
Hubble parameter).  Therefore, since $v_{\rm esc} \approx V_{\rm c}$,
where $V_{\rm c}$ is the circular velocity, we have
\begin{equation}
V_{\rm c}=\sqrt{GM \over R_{\rm halo}} \approx H R_{\rm halo},
\label{eq:vc}
\end{equation}
where $R_{\rm halo}$ is the outer radius of the dark halo. Since the
Hubble expansion decreases monotonically with time and the mass of the
halo after formation is either constant or increases with time, it
follows that a halo that evolves in isolation (\ie, without major
mergers) increases its size and mass.  Assuming, for simplicity, that
the halo mass increases with time as $M(t) \propto t^{\xi}$, from
\eq~(\ref{eq:vc}) we have $R_{\rm halo} \propto H^{-2/3}M(t)^{1/3}
\propto H^{-(2+\xi)/3}$ or, neglecting the cosmological constant,
$R_{\rm halo}(z) / R_{\Delta}=[(1+z_{\rm vir})/(1+z)]^{1+\xi/2}$,
where $R_{\Delta}$ is the virial radius at the time of formation.
According to the numerical results presented in \S~\ref{sec:res} the
dark halo core radius, $r_{\rm s}$, remains constant with time.
Therefore, the concentration parameter, $c_\Delta =R_{\rm halo}/r_{\rm
  s}$, evolves as
\begin{equation}
c_\Delta(z, z_{\rm vir})=c_\Delta(z_{\rm vir})\left({1+z_{\rm vir}\over
    1+z}\right)^{1+{\xi \over 2}}.
\label{eq:cgen}
\end{equation}
Here, we note the result found by R03 that the concentration parameter
$c_\Delta(z_{\rm vir})$ of recently virialised haloes is roughly
constant and independent of the halo mass.  Finally, based on simple
scaling arguments \citep[see][]{Gott:75}, the outer
envelope should have a density profile $\rho \propto M(z)R_{\rm
  halo}(z)^{-3} \propto R^{-6/(2+\xi)}$.

\cite{GunnG:72} find that during secondary infall $\xi =2/3$
(infalling shells do not cross during the infall). But if, after the
collapse, crossing shells start leading to violent relaxation then one
expects the equilibrium halo to be more centrally condensed and the
outer envelope should have a logarithmic slope between $-4$ (violent
relaxation) and $-2.25$ (infall upper limit). We also note that the
assumption that the infall rate is a power law in time is an
oversimplification since isolated haloes will evolve differently
depending on whether they are located in an overdense or underdense
region of the IGM or as a consequence of extended density
correlations. In other words, the density power spectrum introduces
preferred length scales, and local departures from $\Omega_{\rm m}=1$
introduce preferred time scales. This more complicated and realistic
accretion history is reflected in a large scatter of the concentration
parameters of haloes of a given mass, as shown by N-body simulations
\citep[\eg,][]{Bullock:01, WechslerB:02}.  For the purpose of this
section we simply note that assuming $\xi \rightarrow 0$ (\ie,
neglecting the mass accreted after virialisation) we obtain good
agreement with $N$-body simulations for the dependence of the
concentration parameter on the redshift and formation redshift,
$c_{\Delta} \propto (1+z_{\rm vir})/(1+z)$
\citep[\eg,][]{WechslerB:02} and outer profile $\rho \propto r^{-3}$
\citep[\eg,][]{Navarro:97}. Our simple arguments are also in good
agreement with the numerical results of \cite{WechslerB:02, Zhao:03,
  Zhao:03b} that find that the concentration of haloes is about
constant ($c_\delta \sim 4$) at the end of an initial phase of rapid
accretion (\ie, virialisation) and increases later to larger values as
a consequence of a slower secondary phase of accretion (\ie, secondary
infall). In agreement with those numerical studies,
\eq~(\ref{eq:cgen}) implies that at high redshift most haloes have
small concentrations $c_\Delta \sim 3-4$, since they did only recently
virialise.


In CDM cosmologies small mass haloes form earlier. We can use the
results of \S~\ref{ssec:inner} to relate the mass of the halo to its
typical redshift of virialisation: $(1+z_{\rm vir}) \propto M_{\rm
  dm}^{-(n+3)/6}$.  Using the relation $\alpha =(9+3n)/(5+n)$ and
$c_\Delta \propto (1+z_{\rm vir})$ we have $c_\Delta(z=0,M) \propto
M_{\rm dm}^{-\alpha/(9-3\alpha)}$.  In Fig.~\ref{fig:cm} we compare
$c_\Delta$ and $\alpha$ as a function of mass with observations
\citep{Kleyna:03, Magorrian:03,deBlok:03,Dahle:03}. The DM profile
slope $\alpha$ and concentration $c_\Delta$ based on the generalised
DM profile are consistent with the observational data if the
concentration parameter at virialisation is $c_\Delta(z_{\rm vir})
\sim 3-4$ \footnote{Note that the haloes in R03 have $c_\Delta \sim
  6-7$, but they have been analysed some time after virialisation.  A
  40 \% increase of their concentration since virialisation is
  plausible.}. Note that the concentrations shown in the bottom panel
are a rough approximation because they have been derived using scaling
arguments and are valid only qualitatively. At high redshifts the halo
concentrations are about constant and almost independent on the halo
masses, as illustrated by the dashed (at $z=3$) and dotted (at $z=10$)
lines in the bottom panel of Fig.~\ref{fig:cm}.

Finally we note that the mass-dependent profile shape coupled with a
mass-independent concentration parameter at virialisation gives
first-order results in good agreement with a universal profile shape
coupled with the mass-dependent concentration parameter, $c_{\rm
  NFW}$, found by \citep{Bullock:01} and others. This renders the
results presented here more consistent with other results than one
might first have guessed (for a discussion see R03).

\section{Summary and Conclusions}\label{sec:con}
Cosmological simulations of the formation of the first galaxies at
$z>10$ show a good agreement with the observed properties of dSphs in
the Local Group \citep{RicottiGSa:02, RicottiGSb:02,
  Ricotti_pr:03,Tassis:03,Susa:04}. Those simulations, because of the
inclusion of radiative feedback effects, can reproduce the observed
number of luminous galactic satellites and the observed internal
properties of dSphs (\eg, mass-luminosity relation,
metallicity-luminosity relation, mass-to-light ratio, half-light
radius). According to those simulations and similar pure $N$-body
simulations \citep{Ricotti:03}, the DM profiles of dwarf galaxies at
$z=10$ have a central core instead of a steep cusp.

In this paper we have used $N$-body simulations to show that the
central DM core found in dwarf galaxy haloes at $z=10$ is dynamically
stable. At low redshift the halo still has a flat core in agreement
with the observed rotation curves of low-mass disk galaxies and
evidence from the Ursa Minor dSph galaxy. We find that the projected
velocity dispersion profile for a dwarf galaxy with a light
distribution similar to that of the Draco dSph residing in any of the
dark haloes in our simulations resembles the observed profile in the
inner regions of Draco dSph, making these haloes plausible hosts for
the observed dSph galaxies.

In this paper we have evolved the simulated haloes as isolated
systems. While some of the satellites of the Milky Way (\eg, the
Sagittarius dSph) are clearly in the process of being destroyed by the
tidal field of the Milky Way\citep[\eg,][]{Mateo:98}, others (\eg, the
Draco dSph) show little evidence for significant tidal effects in
their inner regions \citep[\eg,][]{Kleyna:02,Klessen:03}. It is
therefore reasonable to assume that their inner haloes have
evolved in near-isolation since the formation of their stellar
populations. The persistence of the central DM core in our simulations
confirms the conclusion of R03 that the mass dependence of the inner
slope of dark matter haloes could arise solely from the power spectrum
of primordial density fluctuations.

The concentrations and slopes of the inner profiles of simulated
haloes are found to be in good agreement with available observations
of dwarf galaxies and clusters at $z=0$. The concentrations as a
function of mass and redshift agree with those found in published
$N$-body simulations in the mass range $10^{12} {\rm M}_\odot \simlt M
\simlt 10^{14}$ M$_\odot$ covered by those simulations. We have also
summarised scaling arguments which demonstrate why the inner slope
depends on the halo mass and why the concentration parameter depends
on the time elapsed since virialisation.

\subsection*{ACKNOWLEDGEMENTS}
MR and MIW are supported by PPARC rolling grants at the Institute of
Astronomy, Cambridge. We would like to thank Jarrod Hurley for running
the $N$-body code on a GRAPE-6 board at the American Museum of Natural
History, New York and Jerry Ostriker for useful discussions. We thank
the referee David Weinberg for comments that improved the quality and
clarity of the paper.

\bibliographystyle{/home/ricotti/Latex/TeX/apj}
\bibliography{/home/ricotti/Latex/TeX/archive}

\label{lastpage}
\end{document}